\documentclass[preprint,5p,times,authoryear,twocolumn]{elsarticle}

\usepackage{booktabs}
\usepackage{mathtools}
\usepackage{setspace} 
\usepackage{etoolbox}
\usepackage{enumitem}
\usepackage{fancyvrb}
\usepackage{amssymb}
\usepackage{subfiles}
\usepackage{verbatim,tcolorbox}
\usepackage{ulem}
\usepackage{pdfpages}
\usepackage{tikz}
\usetikzlibrary{positioning}
\usepackage{array}

\usepackage[sharp]{easylist}

\usepackage[dvipsnames]{xcolor}
\definecolor{ao}{rgb}{0.0, 0.5, 0.0}

\usepackage{url}

\usepackage{breakurl}
\usepackage[unicode,breaklinks]{hyperref}
\hypersetup{
	colorlinks=true,
   	linkcolor=blue,
   	citecolor=blue,
    filecolor=blue,
   	urlcolor=blue}

\newcommand{\furl}[1]{\footnote{\url{#1}}}

\usepackage{rotating}

\usepackage{multirow}
\usepackage{tabularx}
\newcolumntype{L}[1]{>{\raggedright\arraybackslash}p{#1}}
\newcolumntype{C}[1]{>{\centering\arraybackslash}p{#1}}
\newcolumntype{R}[1]{>{\raggedleft\arraybackslash}p{#1}}

\renewcommand{\emph}[1]{\textit{#1}}

\begin{document}

\begin{frontmatter}

\title{FALC: An Artifact-Centric Life Cycle Model and Research Agenda for AI Forensics}

\author[unia]{Adrian Johannes Staeves\corref{cor1}}
\address[unia]{Chair for Cybersecurity, University of Augsburg, 86159 Augsburg, Germany}
\cortext[cor1]{Corresponding author:}
\ead{adrian.staeves@uni-a.de}

\author[rub]{Mayur Rattan Jaisinghani}
\address[rub]{Chair for Security and Privacy in Socio-technical Systems, Ruhr-University Bochum, 44801 Bochum, Germany}
\ead{mayur.jaisinghani@rub.de}

\author[rub]{Rebekah Overdorf}
\ead{rebekah.overdorf@rub.de}

\author[unia]{Frank Breitinger\corref{cor1}}
\ead{frank.breitinger@uni-a.de}
\ead[url]{https://www.FBreitinger.de}

\begin{abstract}
AI systems increasingly appear in criminal investigations as tools, targets, and sources of evidence, yet AI forensics lacks a systematic account of the artifacts they produce: investigators often do not know which traces to look for, and for many traces they do find, no established analysis technique exists.
In this paper we first developed the Forensic AI Life Cycle (FALC) that unlike other life cycle models shifts the focus onto the forensic artifacts created during the life cycle of an AI system. Through FALC we derived a reference list of artifacts that might serve an investigation. 
Second, we collected examples for crimes with AI involvement throughout FALC and questions likely to be asked when investigating these crimes.
Third, we propose a taxonomy of forensic fields of expertise with a focus on the current possibility to analyze AI forensic artifacts using established techniques. 
For investigators, the model provides a systematic basis for identifying which artifacts to look for when an AI system is relevant to a case. For researchers, the classification shows where existing forensic methods already suffice and where new techniques are most urgently needed.
\end{abstract}

\begin{keyword}
AI Forensics, AI life-cycle, artifact-centric life-cycle, AI forensic taxonomy
\end{keyword}

\end{frontmatter}

%%%%%%%%%%%%%
\section{Introduction}\label{sec:introduction}
Due to its recent democratization and versatility, AI can appear in a forensic investigation in various ways. It can be an asset to aid investigators, the target of an attack, or a medium for committing a crime. The digital forensics community has acknowledged the need to investigate AI systems~(e.g., \cite{dragonasForensicAnalysisOpenAIs2024a, jeongLangurTraceForensicAnalysis2025, comeauxExMachinaForensic2026}). What is still missing is a structure that tells an investigator where to look for traces in an arbitrary AI system and that tells researchers which parts of the problem are already solved and which are not.

To fill this gap, we developed a life-cycle model for AI systems which focuses on the artifacts that each stage in the life of an AI system leaves behind. We call it the Forensic AI Life Cycle (FALC). Because of its focus, FALC is both an artifact-centric life-cycle model and an artifact-centric reference model for forensic investigations. FALC provides a systematic way to enumerate questions that can arise during an investigation around an AI system and the artifacts available to answer them. Comparing those questions against the current state of forensic practice yields our second contribution: a classification of the fields of expertise that AI forensics requires into fields where existing methods transfer unchanged, fields where existing methods must be adapted, and fields where no adequate method exists yet. This last group marks where research is most urgently needed. 
In particular, our contributions are:
\begin{itemize}
    \item FALC, an artifact-centric reference model of the AI life cycle designed for forensic investigation rather than for process optimization.
    \item A catalogue of forensic artifacts produced at each stage of the AI life cycle, together with where they can be expected to be found.
    \item A collection of crimes and forensic questions associated with each stage, illustrating how AI can be involved in an offense.
    \item A taxonomy of the forensic fields of expertise required by AI forensics, distinguishing directly applicable fields, fields requiring adaptation, and fields that do not yet exist.
\end{itemize}

The remainder of this paper is structured as follows: 
Sec.~\ref{sec:related} situates our work within related research on AI forensics and AI crime. Sec.~\ref{sec:methodology} describes our methodology. Sec.~\ref{sec:falc} highlights the FALC and catalogues forensic artifacts produced in each stage. Sec.~\ref{sec:crimes} gives, for each stage, examples of crimes and the forensic questions they raise. Sec.~\ref{sec:taxonomy} derives the taxonomy of forensic fields of expertise from those questions. In Sec.~\ref{sec:dis_con} we discuss our findings and conclude this paper.

%%%%%%%%%%%%%%%%%%%%%%%%
\section{Background and Related Work}\label{sec:background}\label{sec:related}

Research into how AI changes digital forensics largely falls into two groups: \textit{AI for forensics}, which applies AI to improve forensic investigations (\cite{dunsinComprehensiveAnalysisRole2024, yinDigitalForensicsAge2025, solankeExplainableDigitalForensics2022}); and \textit{AI forensics}, which asks how investigations change when AI systems are involved in a crime (\cite{schneiderAIForensicsDid2023}). This paper addresses the latter by introducing a forensic-focused, artifact-based AI life cycle model that provides a systematic structure for locating evidence when AI systems are investigated. 

\cite{baggiliFoundingDomainAI2019} laid the conceptual groundwork for AI forensics, defining it as the investigation of ``failures in AI enabled systems.'' This definition notably omits cases where AI can serve as evidence without having failed, such as when it was used as a tool to commit a crime. Their taxonomy is correspondingly organized around what is examined, whereas FALC is organized around where evidence originates. The differences also reflect how the field of AI has changed over the past years and the rapid change in how people interact with and use AI systems. In \ref{app:comp_baggili} we relate FALC to their resulting taxonomy. 

Several papers propose classifications of AI crime (\cite{jeongArtificialIntelligenceSecurity2020, sibaiAICrimesClassification2020, kingArtificialIntelligenceCrime2020, blauthArtificialIntelligenceCrime2022}). While these papers discuss \textit{which AI crimes} can happen, our approach focuses on \textit{which artifacts} are created and available for investigators.

Several works investigated specific AI systems or a group of similar AI systems, determining what can be found in these systems during a forensic investigation. Noteworthy examples are AI companion apps~(\cite{comeauxExMachinaForensic2026}), the ChatGPT mobile app~(\cite{dragonasForensicAnalysisOpenAIs2024a}) and OpenClaw~(\cite{gruberFoundationsAgenticAI2026}). While these are valuable case studies, they can not show the bigger picture of AI forensics. 

Outside of forensics, many AI lifecycle models exist. \cite{wangAILifecycleDatadriven2025} identified and compared 20 of the most important ones. These models serve to structure industry processes or ensure ethical AI, and are therefore not well-suited to be directly applied to AI forensic questions. We use these as a starting point for our life cycle model.

%%%%%%%%%%%%%%%%%%%%%%%
\section{Methodology}\label{sec:methodology}
\begin{figure}[t]
\centering
\begin{tikzpicture}[
  stage/.style={
    draw, rounded corners=3pt, line width=0.4pt,
    fill=teal!8, draw=teal!60!black,
    minimum width=3.1cm, minimum height=0.8cm,
    align=center, font=\small
  },
  band/.style={
    draw, rounded corners=8pt, line width=0.4pt,
    fill=black!3, draw=black!40
  }
]
\node[band, minimum width=7.2cm, minimum height=4.7cm] (gov) at (0,0) {};
\node[font=\small\bfseries, anchor=north] at ([yshift=-0.3cm]gov.north)
  {Governance and Quality Assurance};
\node[font=\scriptsize, anchor=north, text=black!60] at ([yshift=-0.75cm]gov.north)
  {Active before and throughout every stage};

\node[stage] (da) at (-1.7,.6) {Data Acquisition};
\node[stage] (dp) at (1.7,0.6) {Data Preparation};
\node[stage] (md) at (-1.7,-0.55) {Model Development};
\node[stage] (mi) at (1.7,-0.55) {Model Integration};
\node[stage] (dep) at (-1.7,-1.7) {Deployment};
\node[stage] (use) at (1.7,-1.7) {Use};

\end{tikzpicture}
\caption{The FALC model. Governance and quality assurance spans the entire life
cycle. The remaining six stages are deliberately drawn without ordering: in
practice they are interwoven, revisited, and sometimes skipped entirely, so an
investigator should search for artifacts from every stage regardless of the stage in which the suspected crime occurred.}
\label{fig:FALC}
\end{figure}
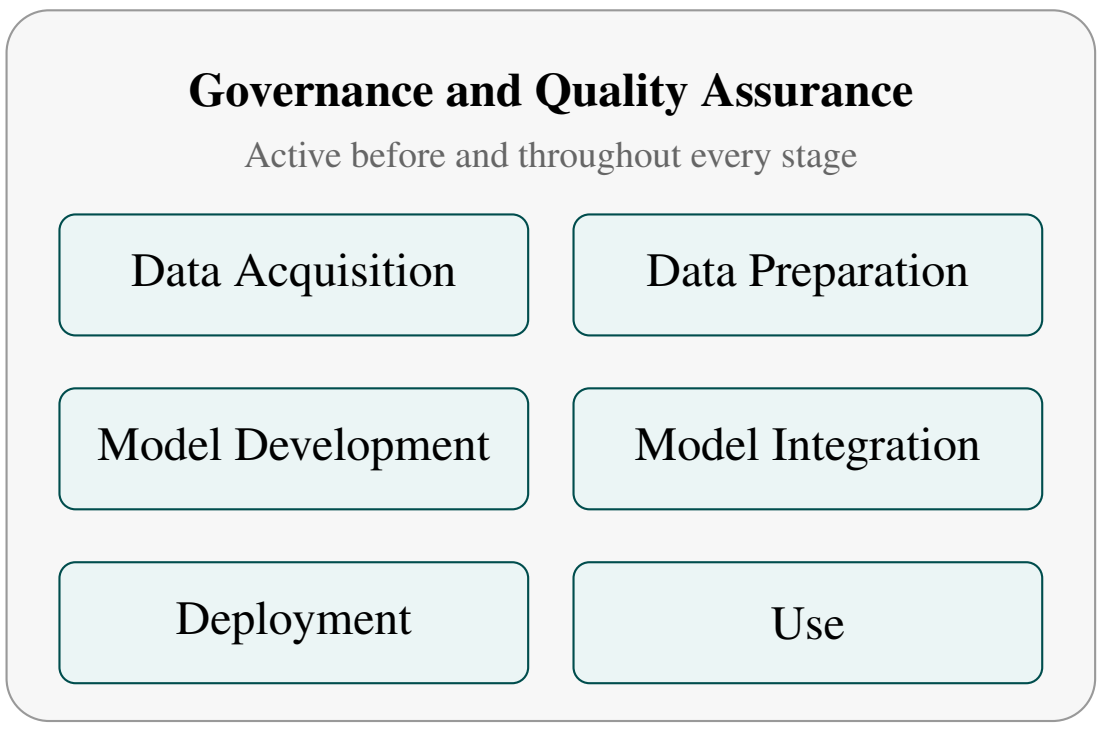

As mentioned in Sec.~\ref{sec:background}, multiple AI life cycle models exist, but none of them is suitable for AI forensics. We therefore developed a new life cycle model that separates stages by the artifacts they produce rather than by when they are performed. The model was derived through conceptual synthesis of existing life cycle models, not a systematic review. 
We first identified common and unique stages and development steps for the 20 most important life cycle models\footnote{TDSP, CRISP-DM, Microsoft model, Allam's AI lifecycle, End-to-end AI lifecycle, AI application project lifecycle, Refined CRISP-DM, AI lifecycle in clinical AI, CDAC AI life cycle, Extended CRISP-DM, Isom's AI lifecycle, Moley's AI lifecycle, Fairness intervention AI lifecycle, AI development lifecycle, AI Ethics Lifecycle, AI-enabled system lifecycle, Edge-to-cloud AI lifecycle, HEAAL, ODD \& BC driven AI lifecycle, Seven-layer model} identified by \cite{wangAILifecycleDatadriven2025}. Next, we determined the forensically relevant artifacts created during each step by reading diverse academic literature describing common practices and challenges faced during each step, looking for direct and implicit mentions of created artifacts\footnote{e.g., It is rarely acknowledged that training a model produces the trained model. Nevertheless, common knowledge implies so.}. 
We deemed the artifact list complete once at least three relevant publications resulted in no change to the list. Finally, we combined the individual steps into stages in a way that each stage creates different types of artifacts. 
Consequently, we arrived at the Forensic AI Life Cycle (FALC) shown in Fig.~\ref{fig:FALC}. In \ref{app:validation} we test FALC by applying it to a real investigation. 

When collecting crimes and forensic questions, our goal was not to create a comprehensive list of all AI-related crimes, but instead to show that crimes can happen during each stage of FALC. Therefore, we decided against a structured literature review and instead searched Google Scholar for AI crime-related literature until we had found at least one example of a crime for each stage of FALC. From these crimes, we derived the questions an investigator may need to answer in such cases. 

Lastly, we created a taxonomy to help AI forensic researchers determine which AI forensic subfields need more research and to help investigators determine which artifacts provide the most value to an investigation. For creating the taxonomy, we followed the empirical-to-conceptual approach described by \cite{nickersonMethodTaxonomyDevelopment2013}.

%%%%%%%%%%%%%%%%%%%%%%%%
\section{FALC: The Forensic AI Life Cycle Model}\label{sec:falc}
Most AI life cycle models differ primarily in how they order stages, and to a lesser extent in how they group the steps needed to build an AI model. Since our goal is to give investigators an overview of which artifacts might exist when examining an AI system, we deliberately leave the stage order undefined. In practice, stages are often interwoven, reordered, or revisited, so investigators should look for artifacts from all stages regardless of a case's specifics.

We considered adding typical connections between stages, but decided against it: doing so could mislead investigators into assuming that a later stage does not need examination if the crime occurred at an earlier one. As artifacts from any stage may be relevant regardless of where the crime took place, we impose no ordering.

Removing the ordering allowed us to regroup the steps of AI system development by the artifacts they produce.
This resulted in the FALC consisting of the stages shown in Fig.~\ref{fig:FALC}.
Not every stage is visited in every case; the model categorizes the steps that \emph{can} occur during the development of an AI system, not those that always do.

In the remainder of this section we briefly describe each of the 7 stages in FALC, and list the artifacts that belong to each stage. We include the list of artifacts, as we reference them in Section~\ref{sec:taxonomy}, and they are an integral part of FALC as a reference model, that can help investigators find all relevant artifacts during an investigation. However, reading the complete list of artifacts is not necessary to understand subsequent sections. 

\subsection{Governance and Quality Assurance}
This stage includes all steps taken to prepare for developing an AI system, such as business understanding, problem design, defining requirements, and creating feasibility studies. 
It also includes all steps taken to ensure the quality of the resulting system, such as documenting, auditing, testing, and monitoring. As defined, this stage is active before and throughout the other stages. All steps of this stage produce documentation and reports that are usually restricted to company internal use. 
Namely, the following artifacts may exist:
\begin{description}
    \item \textbf{Policies and procedures:} All documents that describe how processes are supposed to be done or structured.\footnote{This and subsequent artifact groups that don't cite a source were implicitly derived, usually from multiple sources.}
    \item \textbf{Licenses:} Usually, software use is subject to a license that describes what the user is allowed to do with the software. For bigger projects, different parts can have different licenses.
    \item \textbf{Design documents:} Documents that describe what the system is supposed to do and the problem it is supposed to solve. Examples are the problem statement, requirements, risk assessments, feasibility studies, and ethical evaluations~(\cite{martinez-plumedCASPDMContextAware2017, haakmanAILifecycleModels2021, diazDataPreparationArtificial2021}).
    \item \textbf{Documentation:} Documents that describe how the product works and how the parts work together.
    \item \textbf{Audit results:} Various audits might be performed throughout an AI system's life, such as data quality-, model performance-, risk-, ethical- or law compliance audits~(\cite{liMakingItPossible2025}). The findings of these audits can be of great value to an investigation, as they might confirm the system's state at a specific point in time, hint at potential weaknesses, reveal previously unknown processes, or provide various other insights.
    \item \textbf{Test results:} The results of manual and automated testing can provide insights into the system's functionality at a specific point in time. These include unit, platform, application, and post-deployment tests, as well as model evaluation scores. 
    \item \textbf{Test procedures:} These can be automated test pipelines but also test documentation or instructions for manual testing~(\cite{choudharyEmpiricalStudyVersionControl2026}). The test procedures can give the test results meaning. 
    \item \textbf{Monitoring:} Usually companies monitor their system's performance and user behavior after deployment to get early insights in case of failure or if performance degrades~(\cite{groverENGINEERINGROBUSTAI2025}). Artifacts created through monitoring can give an investigator these same insights. 
    \item \textbf{User tracking:} For reasons such as improving the product or selling the data for advertising, AI systems might track the user's behavior and collect vast amounts of user data~(\cite{comeauxExMachinaForensic2026}). Such data can give investigators additional insights into the users' behavior and motives. Because such data is often collected by or sold to third parties, it may remain obtainable from them even if one party deleted it before the investigation.
    \item \textbf{User feedback:} Companies may solicit feedback directly, or users may post opinions on review sites and social media. Indirect feedback is subjective, hard to locate, and difficult to analyze automatically, but it is usually publicly available and can reveal how user perception and user-facing behavior changed over time.
    \item \textbf{Backups and version history:} As in any software project, backups of the system or its components may exist, and version control software such as git is commonly used~(\cite{choudharyEmpiricalStudyVersionControl2026}). Both reveal the system's state at a given point in time and how it changed during development and use.
\end{description}

\subsection{Data Acquisition}
This stage covers all steps taken to obtain the initial data, which can be acquired in several ways.
\begin{description}
    \item \textbf{First party data:} A company or other entity that collects user data might already have all the information they need. In these cases the data acquisition can be as simple as accessing the own data lake~(\cite{rohSurveyDataCollection2021}).
    
    \item \textbf{Second or third party data:} Many datasets are offered for sale by data brokers~(\cite{chenDataAcquisitionNew2023}). Some datasets are also available for free. If one of these datasets is suitable for the intended use case, one might decide to acquire one or multiple of them. 
    
  \item \textbf{Data generation:} Data can be generated by another AI system or through manual effort such as labeling or surveys~(\cite{rohSurveyDataCollection2021}). It can also be derived from existing sources, for example by scraping image search results or by combining datasets in new ways.
\end{description}

These approaches are often combined. Datasets frequently contain metadata that is unused during training but essential to an investigation, for example, to establish the data's source. Depending on how a dataset was acquired, the following artifacts may be available:
\begin{description}
    \item \textbf{Raw data:} The data that was obtained itself can be essential to an investigation. Note that the raw data can possibly be found at any party involved in the data acquisition.
    
    \item \textbf{Inspection:} Before and/or after the acquisition, the data was probably inspected for its usability. This includes inspecting individual samples as well as metrics that relate to part of or the entire dataset~(\cite{chenDataAcquisitionNew2023}). These metrics may be stored, and the inspections themselves can leave traces on the devices used.
    
    \item \textbf{Networking:} During acquisition, the data was probably transmitted over the internet or a local area network at least once. Therefore, networking-related artifacts might exist.
    
    \item \textbf{Storage:} The data has to be stored. As AI requires large amounts of data, possibly dedicated storage servers might be utilized. 
    
    \item \textbf{Access logs:} If the data was extracted from a data lake or downloaded from a broker's server, traces of the access can be found at the storage server.
    
    \item \textbf{Generation system:} If the data was generated, that system is an artifact of this stage. Among others such an artifact can be a survey system, manual labeling tool or a different AI system that itself might be worth to (partially) inspect.
    
    \item \textbf{Negotiations and applications:} Buying a dataset from a broker or similar source might involve sending an application and/or negotiating the price~(\cite{chenDataAcquisitionNew2023}).
    
    \item \textbf{Monetary Transactions} If the dataset was bought, traces of the monetary transaction might exist. 
\end{description}

\subsection{Data Preparation} 
This stage encompasses all data processing steps taken before training the model. It can take place before data acquisition if preprocessed data is acquired, after the acquisition, or both. In the case of online data augmentation, some data preparation steps can even happen as part of the training pipeline~(\cite{zhouSurveyLLM$times$2025}). It includes data cleaning, data reduction, data preprocessing, and data labeling. What these steps have in common is that they typically process large volumes of data. Depending on when, where, how, and which processing took place, the following artifacts may be available: 
\begin{description}
    \item \textbf{Prepared data:} Creating the prepared data is the main goal of this stage. Because of the high resource intensity of AI systems, new data formats that are optimized for AI might be used~(\cite{zhouSurveyLLM$times$2025})
    
    \item \textbf{Partially prepared data:} Data might be saved after part of the preparation took place. Access to such data can answer questions about the state of the data during different stages of the data preparation. 
    
    \item \textbf{Processing pipelines:} Data preparation usually consists of multiple smaller steps. Pipelines are used to combine these steps~(\cite{zhouSurveyLLM$times$2025}). Access to these pipelines can tell investigators what happened during the data preparation. 
    
    \item \textbf{Collaboration artifacts:} Some data preparation steps such as data labeling are hard to automate and too much effort for one person. Therefore, they are often done crowd-based or performed by a possibly spatially spread-out team~(\cite{rohSurveyDataCollection2021}), which can result in artifacts such as calls for participation, websites as well as network artifacts and collaboration software that was used. If a third-party service was used for the collaboration, some of these artifacts might be found at that third party as well. 
    
    \item \textbf{AI systems:} AI can be used to generate new training data, augment, or label it~(\cite{shortenTextDataAugmentation2021}). If such an AI system was used, analyzing it can help determine whether it introduced errors. 
    
    \item \textbf{Operating system artifacts:} Program execution and file modifications can leave operating system-level traces. These can be found even if other traces of this stage are not available. 
    \item \textbf{Augmentation input} Some data augmentation methods require inputs, such as a knowledge graph for synonym replacements~(\cite{shortenTextDataAugmentation2021}) or a prompt for AI-based data augmentation.
\end{description}

\subsection{Model Development}
This stage covers implementing and training the AI model itself, and is usually the most compute-intensive part of an AI system's life. Fine-tuning and training Low-Rank Adaptations (LoRAs) also belong here. Depending on how training was performed, the following artifacts may be generated:
\begin{description}
    \item \textbf{Final model or adaptation:} The trained or fine-tuned model is the goal. As it will be at the core of the final system, being able to analyze it can greatly benefit an investigation. 
    
    \item \textbf{Training checkpoints:} Because the training is often resource-intensive and might require multiple attempts, intermediate results of the training process might be saved~(\cite{zhouSurveyLLM$times$2025}). These checkpoints can help investigators understand how the model changed during the training process and how it became the way it ended up.
    
    \item \textbf{Base model:} If a model is fine-tuned or adapted in any other way, the base model might be a useful artifact, as it can help understand what the training changed about it.
    
    \item \textbf{Provenance markings:} Some AI companies leave specific markings in the models they train. One way to mark an AI model is to let it memorize an artificial and abnormal data point~(\cite{yangDeepTracerTracingStolen2026}).
    
    \item \textbf{Training parameters:} The success of model training as well as its susceptibility to certain attacks can be heavily influenced by the set of parameters chosen during training~(\cite{decastro-garciaEffectSamplingDataset2019}).
    
    \item \textbf{Training metrics:} As the optimal set of training parameters usually is not known in advance~(\cite{decastro-garciaEffectSamplingDataset2019}), some success metrics such as the validation loss can be tracked during training. These can give an estimate of the model's performance and error rates.
    
    \item \textbf{Training framework and tools:} Training is typically performed using frameworks such as TensorFlow or PyTorch, whose configuration and version can indicate how the model was built. 
    
    \item \textbf{Model code:} Even when using a framework, some code is required to implement the model. This code can help understand how the model is structured and its intended use.
    
    \item \textbf{Dedicated training hardware:} Because training can be highly compute-intensive, it may have been performed on dedicated servers. This can leave traces such as data transfers between compute nodes~(\cite{zhouSurveyLLM$times$2025}), or, if the hardware was rented, contracts, monetary transactions, and usage logs.
    
    \item \textbf{Training optimizations:} To speed up the training, some data might be stored in an optimized file format, or optimizations such as caches can be used~(\cite{zhouSurveyLLM$times$2025}).
\end{description}

\subsection{Model Integration}
An AI Model is not deployed directly and instead integrated into a larger system. Even in cases as ChatGPT, Claude or DALL-E where users think they are directly interacting with the AI model, it is part of a larger system that contains a web interface, input validation, and output post-processing~(\cite{weberLargeLanguageModels2024, shaoAreLLMsCorrectly2025a}). This stage contains all steps taken to integrate the AI model into a larger system.
On a high level, this stage produces the same artifacts as traditional software development, such as source code, compiled programs, default configurations, and databases. Not in this stage belong artifacts that better fit into the \textit{governance and quality assurance} stage. On a more detailed level, the created artifacts are highly diverse and dependent on the purpose of the AI system and its capabilities, which are as diverse as any software can be. This makes it infeasible within the scope of this paper to list them all here. 
Instead, we recommend that investigators approach the artifacts created by this stage by first determining which software components make up the entire software system and then determining the artifacts for each subsystem. These software components can be grouped by when they are executed in relation to the AI model. 
\begin{description}
    \item \textbf{Before the model:} Examples include input generation (e.g., for agentic invocations), input validation, input pre-processing, context enhancement (e.g., for RAG), tokenization, and prompt templates.
    \item \textbf{During model execution:} Examples include tools, memory access, and environmental interactions.
    \item \textbf{After the model:} Examples include output filtering, output conversion, output projectors, and modality generators\footnote{used for generating output in a modality the model was not trained on (e.g., audio files for LLMs)}.
    \item \textbf{Independent of the model:} AI systems can contain software components that operate completely independent of the AI model, e.g., because they provide a different functionality. 
\end{description}

\subsection{Deployment}
Depending on the scale of the AI system, various steps might be necessary to deploy it. Deployment can happen to a server, to the devices of end users, or both. 
Note that tests, which are often seen as part of deploying a software system in FALC, are part of the \textit{governance and quality assurance} stage.
\begin{description}
    \item \textbf{Packaged software:} For deployment, the software gets packaged. Among other options, this can be as containers~(\cite{johnArchitectingAIDeployment2021}) for deployment into the cloud or to serverless architectures or as apps for deployment to end-user devices. 
    
    \item \textbf{Compressed model/software:} To deploy software and models they can be compressed~(\cite{brayfordDeployingAIFrameworks2019}). Access to these compressed versions, before and after they were transmitted to the production server, can help determine whether they have been tampered with. 
    
    \item \textbf{Rewrites and re-implementations:} Due to different requirements in production systems, some parts of an AI system might get re-implemented (e.g., in a different programming language) as part of the deployment~(\cite{johnArchitectingAIDeployment2021}). As this can introduce vulnerabilities, errors, or even backdoors, being able to investigate how the software changed during this stage can be valuable for an investigation. 
    
    \item \textbf{Deployment pipelines:} One goal of MLOps and DevOps is to automate the deployment of the system using automated pipelines~(\cite{gopinathanSoftwareEngineeringPractices2025}). Access to these pipelines can tell investigators what happened during deployment and at what point an adversary might have intercepted it. 
    
    \item \textbf{Hosting agreements:} If the software is deployed to rented servers, a hosting agreement as well as monetary transactions are likely to exist. Sometimes, the AI model itself is hosted separately due to its high computational demands.
    
    \item \textbf{Deployment Configurations:} During deployment, a system connects to the services it relies on by configuring endpoints and provisioning API keys, access tokens, or service accounts, often stored in configuration files, environment variables, or secret managers.  It can help establish a link between the external services on which a system depends and who is responsible for them.
\end{description}
Note that, except for the deployment pipeline, these artifacts might be found at the deploying/distributing entity as well as at the entity the system is distributed to, and differences between the two versions can be essential evidence.

\subsection{Use}
After an AI system is deployed, it can be used. This stage only encompasses the providing of the system and the users using the system. All monitoring steps taken, including the ones to monitor the use, are part of \textit{governance and quality assurance}. Nevertheless, the use of an AI system can produce a vast amount of artifacts that can be of interest. Depending on how the system is set up, these artifacts can be found on the end user's device, on a server used to provide the service, or both. 
\begin{description} 
    \item \textbf{Inputs:} AI systems might store the inputs the user provided even after they were processed. This category also includes other forms of input such as sensor data. 
    
    \item \textbf{Model outputs:} An AI system may store the result of AI model invocations. Examples include chat history with an LLM~(\cite{dragonasForensicAnalysisOpenAIs2024a}) or generated images.
    
    \item \textbf{Intermediate generations:} Some AI systems generate output that is not shown to the user. Examples include the chain of thought from a reasoning LLM or a planned trajectory from an agentic system.
    
    \item \textbf{Provenance markings:} Some AI systems leave specific markings in the content they generate, such as a watermark, C2PA manifests, or generation settings stored in the metadata. 
    
    \item \textbf{Tool implementations:} Tool use means that an AI model, usually an LLM can call certain tools to interact with the environment. These tools can be any program with a predefined API and range from simple data retrieval to file manipulation, entire workflows, and delegating subtasks to a different AI system. The existence of such tools can tell investigators what an AI might have been able to do~(\cite{gruberFoundationsAgenticAI2026}).
    
    \item \textbf{Tool interactions:} When an AI uses a tool, it can leave traces. Besides the expected outcome of the tool use, networking artifacts and log entries might be created~(\cite{gruberFoundationsAgenticAI2026}). These artifacts show that the tool was available to the AI and used.
    
    \item \textbf{Databases and storage files:} Model memory and other persistent data can be found in databases and files that serve a similar purpose~(\cite{alnaqbiForensicInvestigationsAge2025b}).
    
    \item \textbf{Vector databases:} Retrieval Augmented Generation (RAG) and similar methods might use a vector database to store additional data injected during generation~(\cite{zhouSurveyLLM$times$2025}).
    
    \item \textbf{Log files:} Like their non-AI counterparts, AI systems can log certain events with varying amounts of detail~(\cite{alnaqbiForensicInvestigationsAge2025b}).
    
    \item \textbf{Caches:} To prevent the same data from having to be transmitted multiple times, it might be cached and therefore available to an investigation~(\cite{dragonasForensicAnalysisOpenAIs2024a}).
    
    \item \textbf{Configuration files:} They store information about the software's settings~(\cite{gruberFoundationsAgenticAI2026,alnaqbiForensicInvestigationsAge2025b}). While these might not directly be of value to an investigation, they can reveal which other traces might exist.
    
    \item \textbf{Client-server communication:} If the software consists of a client and a server, their communication can serve as artifacts. This includes the network communication, authentication details, and APIs~(\cite{dragonasForensicAnalysisOpenAIs2024a}). These can also be of interest indirectly if the investigators only have access to one of the devices (e.g., the end-user's device), as they can be used to access some data on the other system~(\cite{comeauxExMachinaForensic2026}). 
    
   \item \textbf{Computer use traces:} As with any other application, operating system artifacts can reveal additional information about how the user interacted with an AI system. 
    
\item \textbf{Usage and billing logs:} Models hosted on external infrastructure may have their requests and responses logged, and providers typically retain usage and billing records. Anomalies in these records can serve as early indicators of misuse, and they may be the only evidence available when the model itself is beyond the investigators' reach.
\end{description}

%%%%%%%%%%%%%%%%%%%%%%%%
\section{Crimes and AI Forensic Investigative Questions}\label{sec:crimes}
This section examines each stage of the AI life cycle in turn, giving examples of crimes in which AI is involved at that stage. 
In a second step, we use the crimes to determine AI forensic investigative questions that can be relevant to real cases. We use the term ``AI forensic investigative questions" to describe technical questions directly applicable to a specific AI crime, analogous to digital forensic investigative questions\furl{https://dfiq.org/}. This is to differentiate them from forensic questions, which are often understood as the 5W+1H questions. 

The purpose of this section is twofold. First to illustrate how AI can be involved in a crime and to show that crimes can happen during all stages of AI development. Second to establish a set of AI forensic investigative questions, for use in Section~\ref{sec:taxonomy}. Neither the list of crimes nor the set of AI forensic investigative questions are conclusive, but they span all stages of FALC. Conclusivity is arguably not possible, as AI is still quickly evolving, but also not required for the purposes of this paper. 

\subsection{Governance and Quality Assurance}
Crimes that happen during the \textit{governance and quality assurance} stage tend to impact the AI system only indirectly. An example is purposeful neglect of due diligence, such as not addressing model drift, inadequate testing, or inadequate design leading to harm~(\cite{cummingsTaxonomyAIHazard2024, sibaiAICrimesClassification2020}). Another example is an adversary sabotaging the design of the project, e.g., data poisoning during early planning, leading to harm down the 
line. 
The main AI forensic investigative questions regarding this stage revolve around accountability; either the company's accountability for some failure or what led to a crime and the role certain individuals played in that. Artifacts from this stage can also be useful for answering investigative questions raised by other stages. For example, to narrow down when a certain behavior first occurred. 

\subsection{Data Acquisition}
Illegal data acquisition is one example of a crime in the \textit{data acquisition} stage. It ranges from obtaining data without the required permissions to acquiring material that is illegal to possess, such as CSAM. Here, the investigative questions typically concern how and from where the data was obtained. 

Data poisoning and backdoor attacks~(\cite{hossainReviewAttacksArtificial2024}) are also crimes that belong to the \textit{data acquisition} stage, if the malicious modifications were performed before the data was acquired. Here, the AI forensic investigative questions are likely to revolve around the state of the data at acquisition time and how that affects the resulting model. 

\subsection{Data Preparation}
Data poisoning is a crime that belongs to the \textit{data preparation} stage if the malicious alterations were performed after the dataset was acquired. Here, AI forensic investigative questions are likely to revolve around how the data changed. That includes what the data looked like before and after the transformations, which transformations were performed, and the effects of the transformations. 

\subsection{Model Development}
Training parameters can have a huge impact on the resulting model's performance~(\cite{decastro-garciaEffectSamplingDataset2019}). Making malicious modifications to them is a potential attack to sabotage a model's performance. If such a crime is suspected, the main investigative questions are likely to revolve around whether the training parameters were modified, what impact that had, and who is accountable. 

Another crime we want to discuss here is training a model with a malicious goal, such as aiding in criminal activities. Here, AI forensic investigative questions are likely to revolve around what the model does, why it behaves the way it does, and whether that behavior is intended or not. 

Generative AI systems might exactly reproduce some of their training data. This phenomenon is called memorization and can become a problem if the reproduced training data is copyrighted material. While the exact mechanisms behind memorization are the subject of ongoing research, some court rulings already exist\footnote{e.g. GEMA v OpenAI, GEMA v Sono.ai, Disney v Midjourney (ongoing), Bartz v Anthropic/Concord Music v Anthropic, New York Times v. OpenAI/Microsoft, Like Company v. Google (ongoing)}, with the main question of whether the AI model stored the copyrighted material. 

The goal of model theft is to create a functionally identical copy of a proprietary AI model. The main investigative question is likely to be whether the created model is stolen or was legitimately trained to fulfill a similar goal. Model theft is a particularly interesting example, as two AI models that are in different stages are involved. As a result, FALC has to be applied to both models separately. 

\subsection{Model Integration}
A crime that belongs to the \textit{model integration} stage is the development of malicious-by-design software that has an integrated AI model (\cite{fritschOverviewArtificialIntelligence2022}). An example of this is a computer virus that uses AI to avoid detection. Forensic investigations surrounding such a system are likely similar to ones where the malicious software does not use AI, with some additional questions regarding the roles, capabilities, and functionality of the integrated AI. 

The analysis of the model's integration is also relevant for some crimes that belong to the \textit{use} stage. For example, if AI forensic investigative questions ask for input validation or the capabilities and tools available to the model. 

\subsection{Deployment}
During deployment crimes can happen that target the AI and ones where AI is part of the crime. An example of a crime targeting the AI system is replacing the model during deployment with a malicious one. Here, AI forensic investigative questions likely revolve around whether that replacement took place and which safety mechanisms were used to detect or prevent such a crime. 

An AI system is part of a crime if the system is deployed without the consent of the owner of the hardware the system is deployed to. That can be because the hosting agreement does not allow for running an AI system or because the AI is part of malware. Here investigative questions are likely to revolve around what exactly the system does, what it consists of, how it works, and the targets the software was deployed to. 

\subsection{Use}
Most crimes belong to the \textit{use} stage, as this is where the system provides the functionality that can be abused and where it is most exposed to the public. All crimes in which AI serves as the instrument of the offense, i.e., Jeong's \textit{AI as a tool} category~(\cite{jeongArtificialIntelligenceSecurity2020}), fall into this stage.
Examples include asking an LLM for advice on how to commit the crime, (partially) automating attacks~(\cite{blauthArtificialIntelligenceCrime2022}) or using AI to generate fake news. For these kinds of crimes, AI forensic investigative questions are likely to revolve around the user's actions, interactions with the AI system, and the user's intent. Some of these crimes can also require investigators to detect whether media was generated using an AI system (deepfake detection). 

Just as in all other stages, an AI system can also be attacked. Model inversion attacks~(\cite{hossainReviewAttacksArtificial2024}), model theft, adversarial attacks~(\cite{kurakinAdversarialMachineLearning2016,hossainReviewAttacksArtificial2024}) and prompt injection are examples. Three forensic questions are likely to arise here. The first concerns which sources the AI system drew on and who is accountable for their content. Secondly, investigative questions might revolve around the actions the AI model took. Finally, questions may concern what caused the system to take a particular action, since only this link allows a malicious action by an AI system to be attributed to a person. This type of question will become more relevant with growing complexity and autonomy of AI systems, as with these, the output is no longer directly related to a specific user input.

%%%%%%%%%%%%%%%%%%%%%%%%
\section{Taxonomy}\label{sec:taxonomy}
This section presents a taxonomy of the required forensic fields of expertise. The goal of this taxonomy is twofold. First, the taxonomy can help AI forensic researchers determine which subfields need more research. Second, it can help investigators determine which artifacts to look for, as those that are not possible to obtain or analyze with existing techniques are likely of lower value to an ongoing investigation. We define the analysis of each artifact to contribute to answering a given AI forensic investigative question as its own forensic field. Put differently, we treat each pairing of investigative question and artifact as a different forensic field of expertise. 

For developing the taxonomy, we followed the approach described by \cite{nickersonMethodTaxonomyDevelopment2013}. Considering the mentioned purposes, we decided on the meta-characteristic of current ability to analyze an artifact in a way that it contributes to answering an AI forensic investigative question as part of a forensic investigation. For ending conditions, we decided on the 5 subjective ending conditions: concise, robust, comprehensive, extensible, and explanatory as described in \cite{nickersonMethodTaxonomyDevelopment2013}. As objective ending conditions, we choose:
\begin{itemize}
    \item All possible pairs of artifacts from Sec.~\ref{sec:falc} with AI forensic investigative questions from Sec.~\ref{sec:crimes} have been examined.
    \item At least one object is classified under every characteristic of every dimension~(\cite{nickersonMethodTaxonomyDevelopment2013}).
    \item Every characteristic is unique within its dimension (i.e., there is no characteristic duplication within a dimension)~(\cite{nickersonMethodTaxonomyDevelopment2013}).
\end{itemize}
Due to the large number of artifacts and AI forensic investigative questions determined in Sec.~\ref{sec:falc} and Sec.~\ref{sec:crimes}, respectively, we mainly used an empirical-to-conceptual approach. 

Once all ending conditions were met, our taxonomy consisted of two dimensions. \textit{Ability to acquire} with the characteristics of \textit{does not apply} (e.g., for follow-up questions, where the artifact is expected to be obtained already), obtainable with \textit{established} techniques, and obtainable when \textit{adapting} established techniques. The second dimension, \textit{Ability to analyze}, has the characteristics \textit{impossible}, analyzable with \textit{established} techniques, analyzable when \textit{adapting} established techniques, and \textit{new} techniques required for analysis. 
This taxonomy fulfills all three objective ending conditions, and it is concise, robust, comprehensive, extendible, and explanatory. We argue for concise and robust despite the low number of dimensions, as the dimensions are directly related to the intended purpose. We also consider the taxonomy comprehensive despite the \textit{ability to acquire} dimension not having a characteristic for needing new methods. We did not add this characteristic because it was not required for any of the sample cases we collected. This case also arguably does not exist, as AI systems fundamentally are computer programs that leave artifacts in the same places as non-AI computer programs. However, if an artifact that requires a new technique to acquire is created, discovered, or theorized in the future, adding this characteristic to the taxonomy is a simple extension. 

Table~\ref{tab:taxonomy} shows our final taxonomy. In many cases, an artifact can not be used to answer an AI forensic investigative question because they are completely unrelated. These cases are impossible to analyze for answering the question characteristic and are not included in the table.

\begin{table*}[p]
    \centering
    \begin{tabular}{m{7.4em}|m{8.8em}|m{23.2em}|m{3.1em}|m{3.5em}}
         Stage & Forensic Question & Relevant Artifacts & Obtain & Analysis \\
         \hline
         Governance and Quality Assurance & Determining accountability & policies and procedures, licenses, design documents, audit results, test procedures, test results, monitoring & est & est \\
         Governance and Quality Assurance & First occurrence of abnormal behavior & test results, audit results, user feedback, backups, and version history & est & est \\
         Data Acquisition & Data source and acquisition method & networking, access logs, inspection, generation system, negotiations and applications, monetary transactions & est & est \\
         Data Acquisition & State of the data when acquired  & raw data, inspection, storage, access logs, generation system & est & adapt \\
         Data Preparation & Data modifications & prepared data, partially prepared data, processing pipeline, collaboration artifacts, operating system artifacts & est & adapt \\
         Data Preparation & Data modifications & AI system, augmentation input & adapt & new \\
         Data Acquisition and Data Preparation & Expected impact of data on the resulting model & raw data, inspection, storage, generation system, prepared data, partially prepared data, processing pipeline, collaboration artifacts, AI system, augmentation input & dna & new \\
         Model Development & Modifications of training parameters and their impact &  training parameters, training metrics, training framework/tools, training optimizations, final model or adaptation, training checkpoints & est & new \\
         Model Development & Model's intended purpose and capabilities & final model or adaptation, training checkpoints, training framework/tools, model code & est & new \\
         Model Development & Memorization of content & final model or adaptation, training checkpoints, training parameters, training metrics, training optimizations & est & new \\
         Model Development & Is the model the result of model theft & final model or adaptation, training checkpoints, base model, provenance markings, training parameters, training metrics, training framework/tools, model code, dedicated training hardware, training optimizations & adapt & new \\
         Model Integration & Role, capabilities, and functionality of AI that is part of malware & software, final model or adaptation (\textit{Model Development} stage) & adapt & new \\
         Deployment & Was the model replaced during deployment & deployment pipeline, hosting agreements, packaged software, compressed models/software & est & est\\
         Deployment & Purpose of the software & packaged software, compressed model/software & est & new \\
         Deployment & Target software was deployed to & packaged software, rewrites and re-implementations, deployment pipelines, hosting agreements, deployment configurations & est & adapt \\
         Use & User interactions and intent & inputs, model outputs, intermediate generations, provenance markings, databases and storage files, log files, caches, configuration files, client server communication, computer use traces & est & est \\
         Use & Deep fake detection & model outputs, provenance markings & est & new \\
         Use & Actions performed by AI & model outputs, intermediate generations, tool implementations, tool interactions, databases and storage files, vector databases, log files, caches, configuration files, and client-server communication, computer use traces & est & est \\
         Use & Linking AI system actions to user input & inputs, model outputs, intermediate generations, tool implementations, tool interactions, databases and storage files, vector databases, log files, caches, configuration files, client server communication, computer use traces & est & new \\
         Use & Was this model stolen & inputs, model outputs, intermediate generations, provenance markings, log files, caches, client-server communication & est & new \\
    \end{tabular}
    \caption{Taxonomy of AI forensic investigative questions and artifact pairs. Rows that differ only in the artifact are grouped. Not included are the forensic disciplines where it is fundamentally impossible to contribute to answering the question by analyzing the artifact. \textbf{dna}: does not apply, \textbf{est}: obtainable/analyzable with established techniques, \textbf{adapt}: obtaining/analysis need adaptations to established techniques, \textbf{new}: new analysis techniques are required}
    \label{tab:taxonomy}
\end{table*}

In the following, we describe some challenges and open research questions when obtaining an artifact or analyzing it to answer a specific question, which served as the rationale behind placing them in a different characteristic than obtainable/analyzable using established techniques. These are the forensic fields of expertise where most research is still needed. 

\paragraph{Large amounts of data}
The training of AI models requires large amounts of data. If an investigation requires finding malicious modifications of such data, manually investigating the dataset is not feasible due to its volume, and automated anomaly detection is currently too unreliable to be used. Therefore, we placed the forensic fields that require in-depth analysis of large amounts of training data, such as finding data modifications based on the prepared data of the \textit{data preparation} stage, as needing adaptations to the existing automated anomaly detection mechanisms. 

\paragraph{Impact of training input on model}
A further challenge arises from open questions surrounding model training. Although research has examined how training data, hyperparameters, and model architecture shape the resulting model~(\cite{mohammedEffectsDataQuality2025, liaoEmpiricalStudyImpact2022, raghuVisionTransformersSee2021}), training in practice still relies largely on trying different configurations and selecting the most successful one. As a result, a given training choice cannot reliably be traced back to an intention. Investigators are therefore rarely able to determine from a model's behavior alone whether a particular property was introduced deliberately or emerged incidentally.
As far as we are aware, there are currently no forensic techniques that allow for such predictions, and most other techniques for predicting the outcome of model learning are unreliable. Therefore, we categorized all forensic fields of expertise where the question requires predicting the impact of something on the outcome of the training process as requiring new forensic analysis techniques. 

\paragraph{XAI}
AI forensics faces a similar problem for questions that require understanding how a model reaches its conclusions, such as its intended purpose and capabilities, its reliability, or whether it memorized particular content. The way models are trained makes their internal reasoning difficult to reconstruct, and such understanding is usually not required to operate the system. Explainable AI (XAI) develops methods to determine how and why a model produced a given output, but these methods have not yet been evaluated for their forensic value. Most require valid input data, which may not be available to an investigation, and they produce results with unquantified uncertainty, making them difficult to defend in court. We therefore classify these questions as requiring new analysis techniques.

\paragraph{Deepfake detection}
Finally, deepfake detection remains an active research problem~(\cite{soundaryaDeepfakeDetectionCritical2026, chandraDeepfakeEval2024MultiModalIntheWild2026, croitoruDeepfakeMediaGeneration2026}), indicating that established forensic techniques do not yet solve it adequately. We therefore classify the fields concerned with deepfake detection as requiring new analysis techniques.

%%%%%%%%%%%%%%%%%%%%%%%%
\section{Discussion and Conclusion}\label{sec:dis_con}

Applying the taxonomy across all stages reveals that forensic difficulty is not evenly distributed over the life cycle of an AI system. The stages surrounding the model are largely tractable: governance and quality assurance, data acquisition and deployment mostly produce known artifacts that established methods can acquire and analyze. Difficulty concentrates in the stages closest to the model itself. Data preparation strains existing methods through sheer volume, and model development and use raise questions current techniques cannot answer at all, as they require an understanding of the model's internal behavior that does not yet exist. AI forensics is therefore not uniformly immature: its periphery is well served, and its center is not, which suggests that research effort is best directed at the model rather than at the surrounding infrastructure.

This also shapes how investigations should proceed. Rather than starting from the model, investigators can ask which stages a system passed through and work outwards from those whose artifacts are conventional and relevant to the specific case. \textit{Governance and quality assurance} is usually the most productive entry point: its artifacts are ordinary documents, audit results and version histories; they need no AI-specific techniques, and they frequently date and corroborate findings from other stages. Currently, artifacts such as the model itself are often best treated as a last resort, as they are the ones for which no reliable analysis method exists.

AI systems are usually built by more than one party, with dataset preparation outsourced, and the service hosted on rented infrastructure; the same artifacts are produced and only their location changes. That location determines which jurisdiction and legal process applies, so an artifact may or may not be obtainable depending on who holds it. 

Several AI systems may also be involved in one crime, for instance, when an adversary uses one model to attack another. Where the models remain separate, they can be investigated independently. Where they collaborate or form one system, stages completed before the models were combined are investigated separately, while later stages treat the combined system as a unit: \textit{data acquisition}, \textit{data preparation} and \textit{model development} are examined per model, \textit{model integration} and \textit{use} jointly. \textit{Governance and quality assurance} follows whichever party was accountable for the stage, and \textit{deployment} is treated jointly if the models were deployed together. 

FALC does not feature an end-of-life or decommissioning stage. Decommissioning
does produce artifacts, such as deletion logs, backup expiry records and
certificates of destruction, but the questions they raise are predominantly
anti-forensic ones: whether data claimed to be deleted was in fact deleted, and
whether residual copies remain. As anti-forensics is out of scope for this
paper and warrants treatment of its own, we exclude the stage here.
Investigators should nonetheless bear in mind that artifacts from any stage may
have been deliberately destroyed, and we consider the forensic analysis of
decommissioning a worthwhile direction for future work.

Several limitations constrain our results. FALC was derived conceptually rather than empirically and evaluated against a single case study, selected because it exercises different stages across the life cycle. This demonstrates applicability rather than coverage: the model can be used to structure an investigation, but its artifact catalogue is not shown to be complete for other system types.
The artifact catalogue reflects current practice; new artifacts and possibly new stages will have to be added as AI develops, although the model accommodates new artifact types without restructuring. The crimes we list are illustrative rather than systematically collected. The assignment of fields to categories also involves judgment: a field requiring new domain knowledge could be argued to require adaptation instead, and we draw the line at whether an existing method yields a valid result when applied to the artifact. We expect fields to migrate between categories as adapted techniques become established, making the taxonomy a snapshot whose re-application would show where the field has matured. Finally, our sources are drawn predominantly from English-language literature and industry practice.

\section{Conclusion}
We presented FALC, an artifact-centric life cycle model for AI systems, a catalogue of the forensic artifacts each stage produces, and a taxonomy classifying the forensic fields AI investigations require by whether existing methods apply, need adaptation, or do not yet exist. Applying the taxonomy shows that forensic difficulty is concentrated in the stages closest to the model itself, while the surrounding stages are largely served by established methods. For practitioners, the catalogue offers a structured basis for scoping what to preserve and request in a case involving AI, including artifacts that lie outside the system under examination. Together, these provide a common foundation for AI forensics, but they are only a first step. As AI systems and the techniques to analyze them continue to evolve, we expect both to be refined and extended. We hope this work encourages the forensic and AI communities to jointly build the methods, tools, and practices needed to make all parts of AI systems investigable.

\section*{Disclosure of AI-Assisted Writing Tools}
Several authors used ChatGPT, Claude, and Grammarly to help with tasks such as revising, condensing text, and addressing grammatical errors, typos, and awkward phrasing. All AI-generated suggestions were thoroughly reviewed and adjusted where needed to ensure they accurately reflected the authors’ intended meaning before being incorporated into the paper.
%
%%%%%%%%%%%%%%%%%%%%%%%%
\section*{CRediT Authorship Contribution Statement}
\textbf{Adrian Johannes Staeves:} Conceptualization, Methodology, Validation, Investigation, Writing - Original Draft, Visualization
\textbf{Mayur Rattan Jaisinghani:} Writing - Review \& Editing
\textbf{Rebekah Overdorf:} Writing - Review \& Editing
\textbf{Frank Breitinger:} Conceptualization, Writing - Review \& Editing, Visualization, Supervision
%
%
%%%%%%%%%%%%%%%%%%%%%%%%
\section*{Declaration of Interest}
The authors declare that they have no known competing financial interests or personal relationships that could have appeared to influence the work reported in this paper.

\bibliographystyle{bibl/model5-names}
\bibliography{bibl/literature.bib}

\onecolumn
\appendix

\section{Comparison with \cite{baggiliFoundingDomainAI2019}}\label{app:comp_baggili}

\begin{table}[h]
    \centering
    \begin{tabular}{c|c|c|c|c|c|c|c|c}
         & & \rotatebox{90}{Governance and Quality Assurance} & \rotatebox{90}{Data Acquisition} & \rotatebox{90}{Data Preparation} & \rotatebox{90}{Model Development} & \rotatebox{90}{Model Integration} & \rotatebox{90}{Deployment} & \rotatebox{90}{Use} \\ \hline
         \multirow{3}{*}{AI Training Forensics} & Training Process Forensics & & & & X & & & \\
         & Dataset Forensics & & X & X & & & & \\
         & Environment Forensics & & X & X & & & & \\ \hline
         AI Substrate Forensics & & & X & X & X & & & X \\ \hline
         AI Application Forensics & & & & & & X & & X \\ \hline
         \multirow{5}{*}{AI Model Forensics} & Model Authentication Forensics & & & & X & & & \\
         & Model Identification/Extraction Forensics & & & & X & & & X \\
         & Model Ballistics Forensics & & & & X & & & X \\
         & Model Performance Forensics & & & & X & & & \\
         & Model Malware Forensics &  & & & & & & X \\
    \end{tabular}
    \caption{The subfields defined by \cite{baggiliFoundingDomainAI2019} and where they fit into FALC.}
    \label{tab:comp_baggili}
\end{table}

\clearpage
\section{Applying FALC to the artifacts discovered in NTSB Highway Investigation HWY18MH010}\label{app:validation}
To test the completeness of FALC we applied it to a real investigation of an AI system. We chose the NTSB Highway Investigation HWY18MH010 as it is a case where the investigators determined the responsibility of both the developer of the AI system (Uber ATG) and the person using it (the safety driver), the case gained some popularity, caused regulatory changes and the NTSB docket with 44 items is publicly available. Table~\ref{tab:validation} shows the 30 artifacts mentioned in the docket items and where they fit into FALC. 

\begin{table}[h!]
    \centering
    \begin{tabular}{c|m{19em}|l|l}
         Docket No. & AI related artifact & FALC stage & FALC artifact group \\ \hline
         \multirow{2}{*}{4} & \multirow{2}{*}{Sensors used by automation system} & Governance and Quality Assurance & documentation \\
         & & Model Integration & before the model \\
         4 & Sensor calibration schedule and last calibration & Governance and Quality Assurance & policies and procedures \\
         4 & Procedure to plan a route & Governance and Quality Assurance & policies and procedures \\
         4 & ADS engagement procedure and restrictions & Governance and Quality Assurance & policies and procedures \\
         4 & ADS disengagement procedures and triggers & Governance and Quality Assurance & policies and procedures \\
         4 & Object detection and classification, and path prediction & Governance and Quality Assurance & documentation \\
         4 & Hazard avoidance and emergency breaking procedures & Governance and Quality Assurance & policies and procedures \\
         4 & Table 1 (speed measurements, classifications, and predictions) & Use & inputs, model outputs \\
         4 & Crash history of Uber ADS cars & Governance and Quality Assurance & monitoring \\
         17 & Sensor measurements & Use & inputs \\
         18 & safety division and its culture & Governance and Quality Assurance & policies and procedures \\
         18 & hiring procedures and standards & Governance and Quality Assurance & policies and procedures \\
         18 & Cell phone and no texting policy & Governance and Quality Assurance & policies and procedures \\
         18 & Seat belt policy & Governance and Quality Assurance & policies and procedures \\
         18 & Drug test policy & Governance and Quality Assurance & policies and procedures \\
         18 & Maximum driving time & Governance and Quality Assurance & policies and procedures \\
         18 & Fatigue policy & Governance and Quality Assurance & policies and procedures \\
         18 & Reward and disciplinary programs & Governance and Quality Assurance & policies and procedures \\
         18 & Additional policies & Governance and Quality Assurance & policies and procedures \\
         18 & Vehicle operator training & Governance and Quality Assurance & policies and procedures \\
         18 & Shifts and assigned shift & Governance and Quality Assurance & policies and procedures \\
         18 & ADV testing procedures & Governance and Quality Assurance & policies and procedures \\
         18 & VO's responsibilities & Governance and Quality Assurance & policies and procedures \\
         18 & Post crash internal audit & Governance and Quality Assurance & audit results \\
         18 & Post crash external audit & Governance and Quality Assurance & audit results \\
         29 & Intended Human SDV system interactions & Governance and Quality Assurance & policies and procedures \\
         29 & when the SDS identified the pedestrian and its reactions & Use & model outputs \\
         37 & description of different policies & Governance and Quality Assurance & policies and procedures \\
         42 & description of camera system & Governance and Quality Assurance & documentation \\
         42 & captured video and sensor data & Use & inputs \\
    \end{tabular}
    \caption{AI system related artifacts mentioned in the NTSB Highway Investigation HWY18MH010 Docket and where they fit into FALC. Some artifacts included in this list are only very loosely related to the AI system and its development, yet they are still covered by FALC.}
    \label{tab:validation}
\end{table}

\end{document}